# A new Fragile Points Method (FPM) in computational mechanics, based on the concepts of Point Stiffnesses and Numerical Flux Corrections

Leiting Dong[1, *], Tian Yang[2], Kailei Wang[2], Satya N. Atluri[3]

[1] Professor, School of Aeronautic Science and Engineering, Beihang University, China

[2] Graduate Student, School of Aeronautic Science and Engineering, Beihang University, China

[3] Presidential Chair & University Distinguished Professor of Texas Tech University, USA

* Corresponding author. Email address: ltdong@buaa.edu.cn (L. Dong).

**Abstract** In this paper, a new method, named the Fragile Points Method (FPM), is developed for computer modeling in engineering and sciences. In the FPM, simple, local, polynomial, discontinuous and Point-based trial and test functions are proposed based on randomly scattered points in the problem domain. The local discontinuous polynomial trial and test functions are postulated by using the Generalized Finite Difference method. These functions are only piece-wise continuous over the global domain. By implementing the Point-based trial and test functions into the Galerkin weak form, we define the concept of Point Stiffnesses as the contribution of each Point in the problem domain to the global stiffness matrix. However, due to the discontinuity of trial and test functions in the domain, directly using the Galerkin weak form leads to inconsistency. To resolve this, Numerical Flux Corrections, which are frequently used in Discontinuous Galerkin methods are further employed in the FPM. The resulting global stiffness matrix is symmetric and sparse, which is advantageous for large-scale engineering computations. Several numerical examples of 1D and 2D Poisson equations are given in this paper to demonstrate the high accuracy, robustness and convergence of the FPM. Because of the locality and discontinuity of the Point-based trial and test functions, this method can be easily extended to model extreme problems in mechanics, such as fragility, rupture, fracture, damage, and fragmentation. These extreme problems will be discussed in our future studies.

## 1. Introduction

The Finite Element Method (FEM) (Zienkiewicz, Taylor, & Zhu, 2005) based on interelement-continuous trial and test functions, and Element Stiffness matrices, has been widely employed in diverse engineering fields such as aeronautics, automobiles and construction. For structural and solid mechanics problems, the FEM is the most popular approach, and it is also mature and reliable for the analysis of displacements and stress under normal conditions. Because of the existence of continuous non-overlapping element topologies, and continuous shape functions, the modeling under extreme load conditions involving rupture, fracture, fragility and fragmentation is very difficult in the FEM, without resorting to other expediencies such as cohesive zone models, non-local (Peridynamics) theories, vanishing finite elements, etc. We will show later that in the present Fragile Points Method, fragility, rupture, fracture, and fragmentation can be handled very simply and very naturally within the context of the usual continuum mechanics theories.

To eliminate the drawbacks of the traditional FEM, numerous meshless methods have been invented by various scientists. These meshless methods can be divided into two categories: weak form based ones, and particle based ones, respectively. The Element Free Galerkin (EFG) (Belytschko, Lu, & Gu, 2010) and the Meshless Local Petrov-Galerkin (MLPG) (Atluri & Zhu, 1998) methods are two typical meshless approaches, which are based on the Global Galerkin and Local Petrov-Galerkin weak forms, respectively. Their node-based trial functions are required to be continuous over the entire domain. In order to satisfy this requirement, Moving Least Squares (MLS) or Rational Basis Function (RBF) approximations are commonly utilized. However, if we deduce the trial functions by these two methods, we will severely increase the complexity of trial functions, and therefore it will be very tedious to integrate the weak forms, and to impose the boundary conditions. It is noted that the EFG leads to symmetric matrices, while the MLPG leads to non-symmetric matrices. From the difficulties associated with these meshless weak form methods such as EFG and MLPG, we can see that local, very simple and possibly discontinuous polynomial type of trial functions are more desirable for developing an efficient numerical algorithm.

Smoothed Particle Hydrodynamics (SPH) method is one of the most widely used meshless particle methods. It was first proposed (Lucy, 1977) for astrophysics, then applied to fluid and solid mechanics fields (Libersky, Petschek, Carney, Hipp, & Allahdadi, 1993). Due to its meshless characteristics, the SPH method is used for a variety of extreme problems. But several defects in the SPH method limit its applications. For instance, the SPH method is based on a strong form; so, it is hard to prove the method's stability. On the other hand, if we employ Smoothed Kernel functions in the SPH method to compute the derivatives approximately, it could lead to tensile instabilities. Therefore, we prefer a weak form method and prefer to derive the gradients from the interpolation functions.

Other methods such as the Peridynamics (Silling & Bobaru, 2005) and Discrete Element Method (DEM) (Cundall & Strack, 2008) were designed to simulate extreme load problems. In the Peridynamics method, the system is derived by integral equations instead of differential ones. In the DEM, we assume that materials are composed of discrete and separate particles. Both methods are no longer based on the classical continuum mechanics theories; and hence numerous accumulated engineering experiences and constitutive models are discarded. On the other hand, it can be seen that numerical algorithms based on the continuum mechanics can be employed more widely in diverse fields.

From the above analysis, we can note that a general method, based on continuum mechanics, which is able to simulate extreme problems involving rupture, fracture and fragmentation; while having local, very simple and discontinuous polynomial trial and test functions is still nonexistent in literature. In this paper, the Fragile Points Method (FPM) is newly introduced to satisfy these requirements.

In our discussion, we consider one- and two-dimensional Poisson equations as the model problems, since Poisson equations are very common in engineering such as in acoustics, torsion and fluid mechanics. To establish trial and test functions, which are local and very simple polynomials based on Points, it will be hard for them to satisfy the global continuity requirement. Therefore, in the FPM, we no longer require the trial and test functions to be continuous. Instead, they are postulated by using the Generalized Finite Difference method, and are discontinuous over the entire domain. Because of the discontinuity of trial and test functions, if we directly use them in the Galerkin weak form, the FPM will be inconsistent. To remedy this, Numerical Flux Corrections, which are widely employed in Discontinuous Galerkin methods (Arnold, Brezzi, Cockburn, & Marini, 2002) are adopted in the present FPM. After these Flux Corrections, a sparse and symmetric global stiffness matrix can be obtained, as a sum of Point Stiffness Matrices. The integrations leading to Point Stiffness Matrices are very trivial. The application of the FPM to problems posed by Poisson equations is discussed in the following sections and several numerical examples are given to demonstrate the high accuracy, robustness, consistency and convergence of the FPM. Due to the locality and discontinuity of trial and test functions, problems involving rupture, fragility, fragmentation and fracture can be simulated easily in the FPM. Therefore, the FPM has great potential to model extreme problems, which will be further discussed in our forthcoming papers.

In the following discussions, we introduce the process of constructing trial and test functions and the concept of Point Stiffnesses in Section 2. Numerical Flux Corrections, and the numerical implementation of the FPM are discussed in Section 3. In Section 4, numerical examples for 1D and 2D Poisson equations are given. This paper ends with a conclusion, and some discussions for further studies are given in Section 5.

## 2.  Local, Polynomial, Point-based Discontinuous trial and test functions; and the concept of Point Stiffnesses

### 2.1 The construction of local simple discontinuous polynomial Point-based trial and test functions

In this paper, we take Poisson equations to be the model problems, as defined in

Eq. (2.1).

$$\begin{cases} -\nabla^2 u = f & \text{in } \Omega \\ u = g_D & \text{at } \Gamma_D \\ \nabla u \cdot \mathbf{n} = \mathbf{g} \cdot \mathbf{n} = g_N & \text{at } \Gamma_N \end{cases} \qquad (2.1)$$

where $\Omega$ is the entire domain; Boundaries $\Gamma_D$ (Dirichlet) and $\Gamma_N$ (Neumann) satisfy that $\Gamma_D \bigcup \Gamma_N = \partial\Omega$, $\Gamma_D \bigcap \Gamma_N = \varnothing$; $\mathbf{n}$ is the unit vector outward to $\partial\Omega$.

Several Points are scattered inside the domain $\Omega$ or on its boundary $\partial\Omega$ (shown in Figure 1(a)). With these Points, the domain can be partitioned into conforming non-overlapping subdomains, and in each subdomain only one Point is contained within (shown in Figure 1(b)). Plenty of approaches are available for the partition of the domain; in this paper, we employ the simple Voronoi Diagram method. Notice that no Element or Mesh topologies exist in the domain; instead, a set of randomly distributed Points is used for the discretization. Then the simple, local, discontinuous polynomial trial and test functions are established based only on those Points.

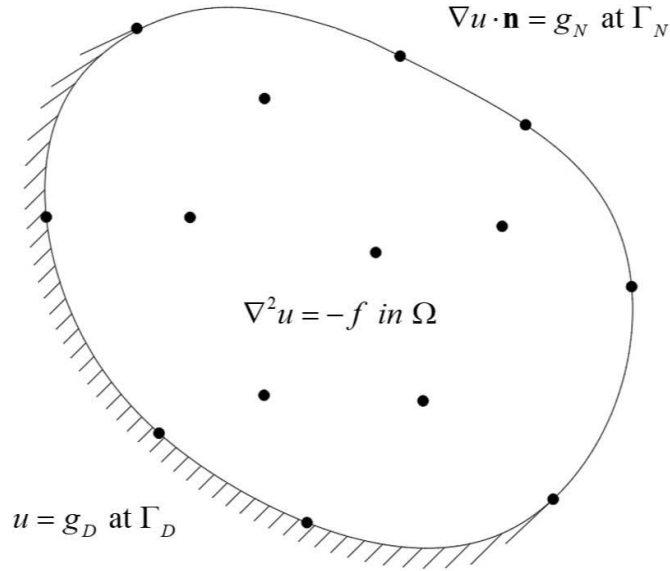

$$\nabla u \cdot \mathbf{n} = g_N \text{ at } \Gamma_N$$

$$\nabla^2 u = -f \text{ in } \Omega$$

$$u = g_D \text{ at } \Gamma_D$$

(a)

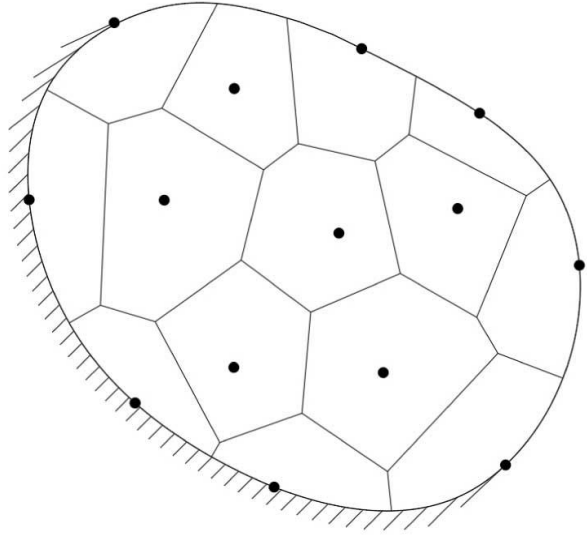

(b)

Figure 1 (a)(b). The domain $\Omega$ and its partition

In each subdomain, the simple local discontinuous polynomial trial function can be defined in terms of the values of $u$ and $\dfrac{\partial u}{\partial x}$, $\dfrac{\partial u}{\partial y}$ at its internal Point. We take 2D linear functions for example and simplicity. Considering the Subdomain $E_0$ which contains the Point $P_0$, the trial function $u_h$ in $E_0$ can be written simply as

$$u_h\left(x,y\right)=u_0+\left.\frac{\partial u}{\partial x}\right|_{P_0}\left(x-x_0\right)+\left.\frac{\partial u}{\partial y}\right|_{P_0}\left(y-y_0\right),\quad\left(x,y\right)\in E_0 \tag{2.2}$$

where $\left(x_0,y_0\right)$ denotes the coordinates of the point $P_0$; $u_0$ is the value of $u_h$ at $P_0$.

Derivatives $\dfrac{\partial u}{\partial x}$, $\dfrac{\partial u}{\partial y}$ at $P_0$ are the unknown coefficients in Eq. (2.2). Determining these two derivatives at each Point is the essential part of constructing local trial and test functions. In this paper, the Generalized Finite Difference (GFD) method (Liszka & Orkisz, 1980) is employed to compute $\dfrac{\partial u}{\partial x}$, $\dfrac{\partial u}{\partial y}$ at $P_0$, in terms of the values of $u_h$ at a number of neighboring Points

To implement the GFD method, we need to define the support of the Point $P_0$ first. The common way is to define a support by taking a circle at the point $P_0$, and all the Points included in that circle are considered to interact with $P_0$ (shown in Figure 2(a)). However, in this paper, we define that the support of $P_0$ involves all the nearest neighboring points for $P_0$ in the Voronoi partition (shown in Figure 2(b)). We name these points as $P_1$, $P_2$,$\cdots$, $P_m$.

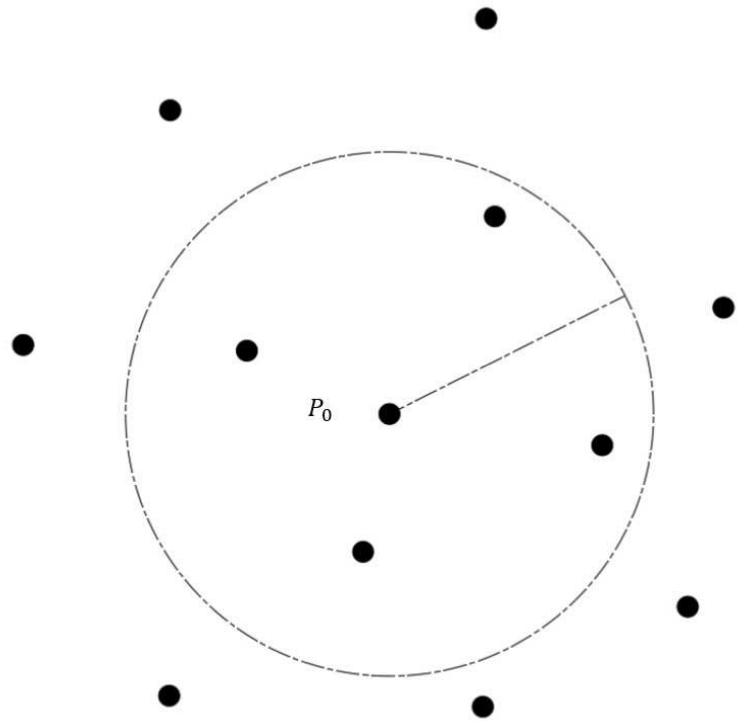

(a)

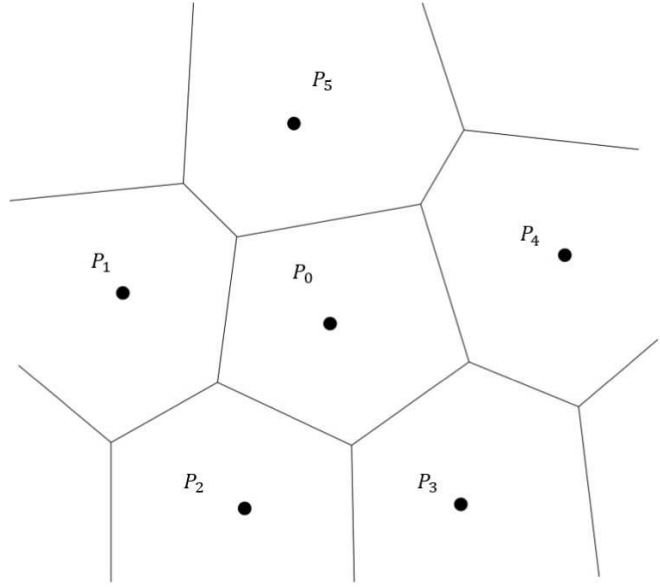

(b)

Figure 2 (a) (b). Two ways of defining the support of $P_0$

To solve the derivatives $\dfrac{\partial u}{\partial x}$ and $\dfrac{\partial u}{\partial y}$ at $P_0$, a weighted discrete $L^2$ norm $J$ is defined as,

$$J = \sum_{i=1}^{n} \left( \left.\frac{\partial u}{\partial x}\right|_{P_0} (x_i - x_0) + \left.\frac{\partial u}{\partial y}\right|_{P_0} (y_i - y_0) - (u_i - u_0) \right)^2 w_i \qquad (2.3)$$

where $(x_i, y_i)$ are the coordinates of $P_i$; $u_i$ is the value of $u_h$ at $P_i$; $w_i$ is the value of the weight function of $P_0$ at $P_i$ $(i = 1, 2, \cdots, m)$. In this paper, we simply use constant weight functions.

The stationarity of $J$ leads to the following formula for the gradient $\nabla u$ at $P_0$.

$$\nabla u = \left( \mathbf{A}^{\mathrm{T}} \mathbf{A} \right)^{-1} \mathbf{A}^{\mathrm{T}} \left( \mathbf{u}_m - u_0 \mathbf{I}_m \right) \text{ at Point } P_0 \qquad (2.4)$$

where

$$\mathbf{A} = \begin{bmatrix} x_1 - x_0 & y_1 - y_0 \\ x_2 - x_0 & y_2 - y_0 \\ \cdots & \cdots \\ x_m - x_0 & y_m - y_0 \end{bmatrix}$$

$$\mathbf{u}_m = \begin{bmatrix} u_1 & u_2 & \cdots & u_m \end{bmatrix}^{\mathrm{T}}$$

$$\nabla u = \begin{bmatrix} \dfrac{\partial u}{\partial x} \\ \dfrac{\partial u}{\partial y} \end{bmatrix}$$

$$\mathbf{I}_m = \left( \begin{bmatrix} 1 & 1 & \cdots & 1 \end{bmatrix}_{1 \times m} \right)^{\mathrm{T}}$$

For convenience, we rewrite Eq. (2.4) with respect to the vector $\mathbf{u}_E$

$$\nabla u = \mathbf{B}\mathbf{u}_E \text{ at Point } P_0 \tag{2.5}$$

where the matrix $\mathbf{B}$ and vector $\mathbf{u}_E$ are defined as below.

$$\mathbf{B} = \left( \mathbf{A}^{\mathrm{T}}\mathbf{A} \right)^{-1} \mathbf{A}^{\mathrm{T}} \begin{bmatrix} -1 & 1 & 0 & \cdots & 0 \\ -1 & 0 & 1 & \ddots & \vdots \\ \vdots & \vdots & \ddots & \ddots & 0 \\ -1 & 0 & \cdots & 0 & 1 \end{bmatrix}_{m \times (m+1)}$$

$$\mathbf{u}_E = \begin{bmatrix} u_0 & u_1 & \cdots & u_m \end{bmatrix}^{\mathrm{T}}$$

Eventually, by substituting the formula of $\nabla u$ at Point $P_0$ into Eq. (2.2), we can obtain the relation between $u_h$ and $\mathbf{u}_E$ as in Eq. (2.6),

$$u_h = \mathbf{N}\mathbf{u}_E, \ \forall (x, y) \in E_0 \tag{2.6}$$

where

$$\mathbf{N} = \begin{bmatrix} x - x_0 & y - y_0 \end{bmatrix} \mathbf{B} + \begin{bmatrix} 1 & 0 & \cdots & 0 \end{bmatrix}_{1 \times (m+1)}$$

The matrix $\mathbf{N}$ is called the shape function of $u_h$ in $E_0$ in terms of $P_0, P_1, P_2, \cdots, P_m$. Since no continuous requirement exists on these internal boundaries, two neighboring subdomains possess their own function values at the common boundary. Therefore, the shape functions are discontinuous at the internal boundaries. The graphs of the shape functions about Point 6 in 1D (11 Points in total) and Point 25 in 2D (49 Points in total) are showed in Figure 3(a) and (b), respectively.

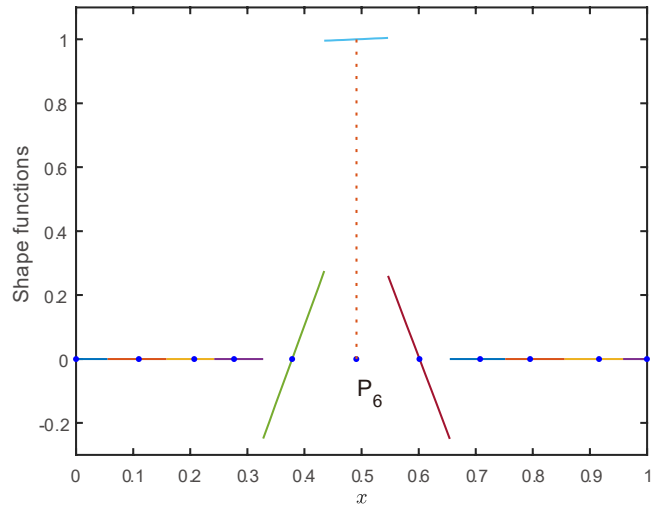

(a)

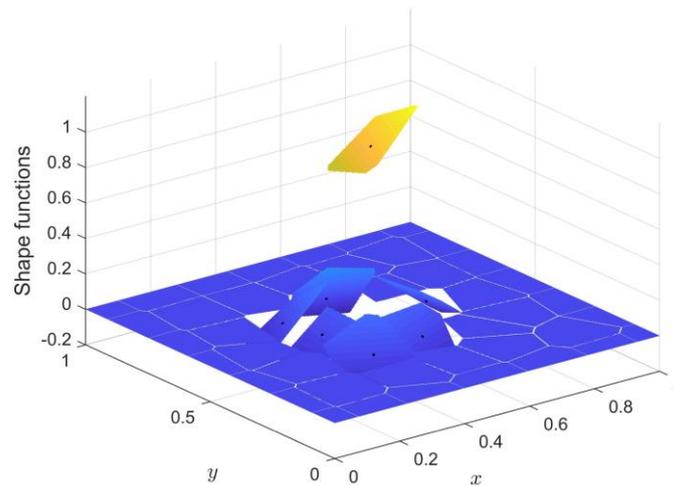

(b)

Figure 3 (a). The shape function for Point 6 in 1D

(b). The shape function for Point 25 in 2D

For every subdomain $E_i \in \Omega$, $u_h$ in $E_i$ can be deduced by the same process. Finally, we can obtain the formula of $u_h$ in the entire domain $\Omega$. As for the test function $v_h$, we prescribe that it has the same shape as $u_h$ in the Galerkin weak form. From the construction of trial and test functions, we can easily see that they are local, simple polynomial, Point-based and piecewise-continuous in $\Omega$. The graphs of trial functions simulating $4^{th}$ order polynomials in 1D (11 points) and 2D (49 points) are shown in Figure 4(a) and (b), respectively.

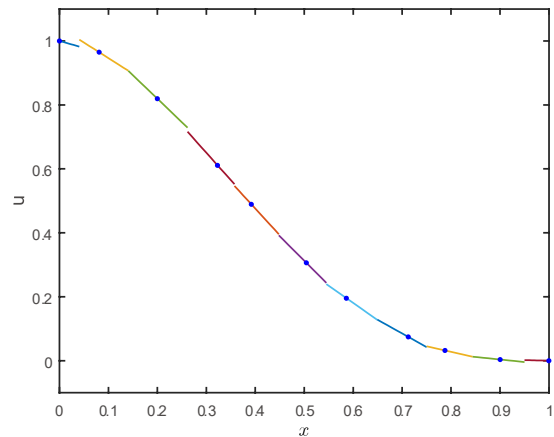

(a)

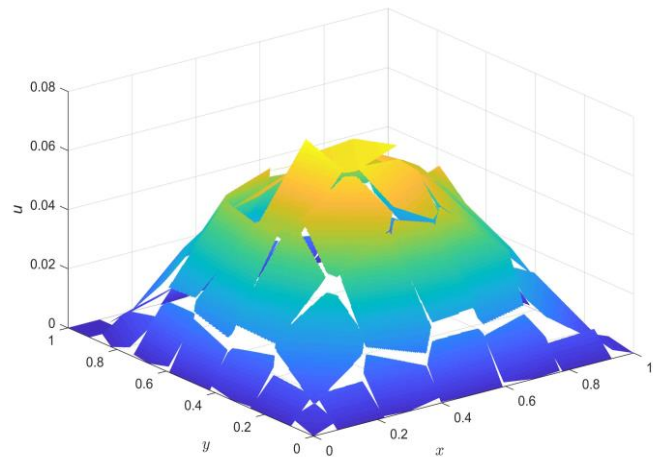

(b)

Figure 4 (a). The trial function in 1D (11 Points)

(b). The trial function in 2D (49 Points)

## 2.2 The concept of Point Stiffnesses

For the Poisson equation, we multiply a test function $v$ on both sides and integrate by parts over the entire domain, then Eq. (2.7) is obtained (omitting boundary terms),

$$\sum \int_{E_i} \nabla u \cdot \nabla v \, d\Omega = \int_{\Omega} f v \, d\Omega \tag{2.7}$$

which is the Galerkin weak form of Eq. (2.1).

We directly substitute the trial and test functions into Eq. (2.7), and only consider the subdomain $E_0$. The stiffness matrix of $E_0$ can be deduced as,

$$\mathbf{K}_{E_0} = \int_{E_0} \mathbf{B}^{\mathrm{T}} \mathbf{B} d\Omega \tag{2.8}$$

where $\mathbf{B}$ is evaluated only at $P_0$. Because we have employed linear interpolations for $u_h$ and $v_h$, $\mathbf{B}$ is a constant matrix in Eq. (2.8). The integration can be directly computed by simply multiplying the value of $\mathbf{B}^{\mathrm{T}} \mathbf{B}$ by the area of $E_0$,

$$\mathbf{K}_{E_0} = \mathbf{B}^{\mathrm{T}} \mathbf{B} S_{E_0} \tag{2.9}$$

where $S_{E_0}$ denotes the area of $E_0$. The global stiffness matrix of the entire domain $\Omega$ can be generated by assembling all the submatrices for each Point, which is the same procedure in the usual Finite Element Method (FEM), except that the FEM involves Element Stiffness matrices, whereas the present Fragile Points Method involves only Point Stiffness matrices. The Point Stiffness matrices can be derived explicitly, as shown in Eq. (2.9). This can also be considered as a one-point quadrature rule as used for linear triangular elements. Furthermore, it can be seen that the present FPM is easily amenable for parallel computations.

Unfortunately, due to the discontinuity of trial and test functions, directly using the Galerkin weak form can result in inconsistency and inaccuracy; in other words, it cannot pass the patch tests. We take the tension of a bar in one dimension for example, which is defined in Eq. (2.10) with the postulated exact solution: $u = x$. Thus, the problem to be solved by the present FPM is taken to be

$$\frac{d^2 u}{dx^2} = 0, \; x \in (0,1)$$
$$u_{x=0} = 0, \; \left. \frac{du}{dx} \right|_{x=1} = 1 \tag{2.10}$$

Linear trial and test functions are employed and 101 Points are distributed in the domain. The errors between the numerical solution and the exact solution of $u$ with uniformly distributed, or randomly distributed Points are shown in Figure 5(a) and (b), respectively.

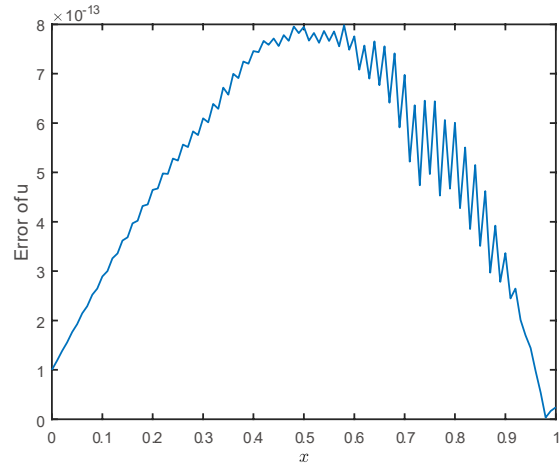

(a)

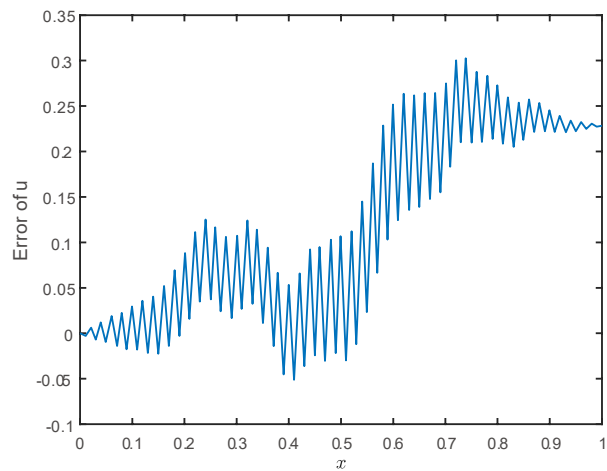

(b)

Figure 5(a). The error of $u$ (101 uniformly distributed Points)

(b). The error of $u$ (101 randomly distributed Points)

It is clear that when Points are distributed randomly, the FPM in the Galerkin weak form is not consistent. To resolve this issue, Numerical Flux Corrections are employed in the present FPM.

## 3. Numerical Flux Corrections

### 3.1 The concept of Numerical Fluxes

In this section, we will introduce the concept of Numerical Fluxes, which are usually used in Discontinuous Galerkin Finite Element Methods (DGFEM) (Arnold, et al. 2002). We rewrite the model problem (Eq. (2.1)) in a mixed form, in other words, using two independent variables: $u$ and $\boldsymbol{\sigma}$, governed by the equations:

$$\begin{cases} \boldsymbol{\sigma} = \nabla u & \text{in } \Omega \\ -\nabla \cdot \boldsymbol{\sigma} = f & \text{in } \Omega \\ u = g_D & \text{on } \Gamma_D \\ \boldsymbol{\sigma} \cdot \mathbf{n} = g_N & \text{on } \Gamma_N \end{cases} \tag{3.1}$$

We multiply the first and second equations in Eq. (3.1) by the test functions $\boldsymbol{\tau}, v$, respectively, then integrate on the subdomain $E$ by parts, resulting in the weak forms:

$$\int_E \boldsymbol{\sigma}_h \cdot \boldsymbol{\tau} d\Omega = -\int_E \mathbf{u}_h \nabla \cdot \boldsymbol{\tau} d\Omega + \int_{\partial E} \hat{u}_h \mathbf{n} \cdot \boldsymbol{\tau} d\Gamma \tag{3.2}$$

$$\int_E \boldsymbol{\sigma}_h \cdot \nabla v d\Omega = \int_E f v d\Omega + \int_{\partial E} \hat{\boldsymbol{\sigma}}_h \cdot \mathbf{n} v d\Gamma \tag{3.3}$$

where $\hat{u}_h$ and $\hat{\boldsymbol{\sigma}}_h$ are called Numerical Fluxes, which are the approximations of $u_h$ and $\boldsymbol{\sigma}_h$ on the boundary of $E$. Numerical solutions $u_h$, $\boldsymbol{\sigma}_h$ should satisfy Eqs. (3.2) and (3.3) for all $E \in \Omega$.

By summing Eqs. (3.2) and (3.3) over all subdomains, we have

$$\int_\Omega \boldsymbol{\sigma}_h \cdot \boldsymbol{\tau} d\Omega = -\int_\Omega u_h \nabla \cdot \boldsymbol{\tau} d\Omega + \sum_{E \in \Omega} \int_{\partial E} \hat{u}_h \mathbf{n} \cdot \boldsymbol{\tau} d\Gamma \tag{3.4}$$

$$\int_\Omega \boldsymbol{\sigma}_h \cdot \nabla v d\Omega = \int_\Omega f v d\Omega + \sum_{E \in \Omega} \int_{\partial E} \hat{\boldsymbol{\sigma}}_h \cdot \mathbf{n} v d\Gamma \tag{3.5}$$

To simplify Eqs. (3.4) and (3.5), we rewrite them using the following notations:

$$\int_\Omega \boldsymbol{\sigma}_h \cdot \boldsymbol{\tau} d\Omega = -\int_\Omega u_h \nabla \cdot \boldsymbol{\tau} d\Omega + \int_\Gamma \hat{u} \cdot \{\boldsymbol{\tau}\} d\Gamma + \int_{\Gamma_h} \{\hat{u}\} \ \boldsymbol{\tau} \ d\Gamma \tag{3.6}$$

$$\int_\Omega \boldsymbol{\sigma}_h \cdot \nabla v d\Omega - \int_\Gamma \{\hat{\boldsymbol{\sigma}}\} \cdot \ v \ d\Gamma - \int_{\Gamma_h} \hat{\boldsymbol{\sigma}} \ \{v\} d\Gamma = \int_\Omega f v d\Omega \tag{3.7}$$

where $\Gamma$ is the set of all external and internal boundaries; $\Gamma_h = \Gamma - \Gamma_D - \Gamma_N$, the internal boundaries. The average operator $\{\}$, and the jump operator are defined respectively as

When $e \in \Gamma_h$ (assuming $e$ is shared by subdomains $E_1$ and $E_2$)

$$\{v\} = \frac{1}{2}(v_1 + v_2), \quad v = v_1 \mathbf{n}_1 + v_2 \mathbf{n}_2$$

$$\{\boldsymbol{\tau}\} = \frac{1}{2}(\boldsymbol{\tau}_1 + \boldsymbol{\tau}_2), \quad \boldsymbol{\tau} = \boldsymbol{\tau}_1 \cdot \mathbf{n}_1 + \boldsymbol{\tau}_2 \cdot \mathbf{n}_2$$

When $e \in \partial \Omega$

$$\{v\} = v, \quad v = v\mathbf{n}, \quad \{\boldsymbol{\tau}\} = \boldsymbol{\tau}, \quad \tau = \boldsymbol{\tau} \cdot \mathbf{n}$$

Different kinds of Numerical Fluxes have been invented and analyzed by various authors (see the review by Arnold, et al, 2002). The choice of Numerical Fluxes can influence the stability, accuracy, consistency and computational time of a method.

## 3.2 Fragile Points Method (FPM)-Primal and FPM-Mixed Approaches

In our current work, for the choice of Numerical Fluxes, we prefer to use Interior Penalty (IP) and Local Discontinuous Galerkin (LDG) Numerical Fluxes (shown in Table 3.1 and 3.2, respectively). With these two Numerical Fluxes, methods are symmetric, consistent and stable (Arnold, et al, 2002).

Table 3.1 The IP Numerical Flux

|  | $\Gamma_h$ | $\Gamma_D$ | $\Gamma_N$ |
|---|---|---|---|
| $\hat{u}_h$ | $\{u_h\}$ | $g_D$ | $u_h$ |
| $\hat{\boldsymbol{\sigma}}_h$ | $\{\nabla u_h\} - \dfrac{\eta}{h_e} u_h$ | $\nabla u_h - \dfrac{\eta}{h_e}(u_h - g_D)$ | $\mathbf{g}$ |

Table 3.2 The LDG Numerical Flux

|  | $\Gamma_h$ | $\Gamma_D$ | $\Gamma_N$ |
|---|---|---|---|
| $\hat{u}_h$ | $\{u_h\} - \boldsymbol{\beta} \cdot u_h$ | $g_D$ | $u_h$ |
| $\hat{\boldsymbol{\sigma}}_h$ | $\{\boldsymbol{\sigma}_h\} + \boldsymbol{\beta} \cdot \boldsymbol{\sigma}_h - \dfrac{\eta}{h_e} u_h$ | $\boldsymbol{\sigma}_h - \dfrac{\eta}{h_e}(u_h - g_D)$ | $\mathbf{g}$ |

In Tables (3.1) and (3.2), $h_e$ is an edge dependent parameter; $\eta$ is a positive number independent of the edge size; vector $\boldsymbol{\beta}$ is a constant on each edge. Notice that the IP Numerical Fluxes are stable only when the penalty parameter $\eta$ is large enough, but the LDG ones only require $\eta > 0$.

We can also find that the IP Numerical Fluxes are only related to the variable $u_h$.

By substituting the IP Numerical Fluxes into Eqs. (3.6), (3.7) and eliminating the variable $\boldsymbol{\sigma}_h$, we can obtain the formula of the FPM with IP Numerical Fluxes only with respect to $u_h$, which is called the FPM-Primal method.

$$
\begin{aligned}
\sum_{E \in \Omega} \int_E \nabla u_h \cdot \nabla v d\Omega &- \sum_{e \in \Gamma_h \cup \Gamma_D} \int_e \left( \{\nabla u\} \; v + \{\nabla v\} \; u \; \right) d\Gamma + \sum_{e \in \Gamma_h \cup \Gamma_D} \frac{\eta}{h_e} \int_e u \; v \; d\Gamma \\
&= \int_\Omega f v d\Omega + \sum_{e \in \Gamma_D} \int_e \left( \frac{\eta}{h_e} v - \nabla v \cdot \mathbf{n} \right) g_D d\Gamma + \sum_{e \in \Gamma_N} \int_e v g_N d\Gamma
\end{aligned}
\tag{3.8}
$$

Notice that Eq. (3.8) consists of one volume integral and some boundary integrals on the left side. The volume integral is just the Point Stiffness we defined in Section 2, and the boundary integrals are the contributions of the Numerical Flux Corrections.

As for the LDG Numerical Fluxes, because the LDG ones are related to both $\boldsymbol{\sigma}_h$ and $u_h$, it is not convenient to eliminate the variable $\boldsymbol{\sigma}_h$ and write the equations in a primal way. Substituting the LDG Numerical Fluxes into Eqs. (3.6) and (3.7), the FPM with LDG Numerical Fluxes are obtained in Eqs. (3.9) and (3.10). We call this method the FPM-Mixed method.

$$
a(\boldsymbol{\sigma}_h, \boldsymbol{\tau}) + b(u_h, \boldsymbol{\tau}) = \int_{\Gamma_D} g_D \boldsymbol{\tau} \cdot \mathbf{n} d\Gamma
$$

$$
b(v, \boldsymbol{\sigma}_h) + J(u_h, v) = -\int_\Omega f v d\Omega - \int_{\Gamma_D} \frac{\eta}{h_e} g_D v d\Gamma - \int_{\Gamma_N} \mathbf{g} \cdot \mathbf{n} v d\Gamma
\tag{3.9}
$$

where

$$
\begin{aligned}
a(\boldsymbol{\sigma}_h, \boldsymbol{\tau}) &= \int_\Omega \boldsymbol{\sigma}_h \cdot \boldsymbol{\tau} d\Omega \\
b(u_h, \boldsymbol{\tau}) &= -\sum_{E \in \Omega} \int_E \nabla u_h \cdot \boldsymbol{\tau} d\Omega + \sum_{e \in \Gamma_h} \int_e u_h \cdot \left( \{\boldsymbol{\tau}\} - \boldsymbol{\beta} \; \boldsymbol{\tau} \; \right) d\Gamma + \int_{\Gamma_D} u_h \boldsymbol{\tau} \cdot \mathbf{n} d\Gamma \\
b(v, \boldsymbol{\sigma}_h) &= -\sum_{E \in \Omega} \int_E \boldsymbol{\sigma}_h \cdot \nabla v d\Omega + \sum_e \int_e v \cdot \left( \{\boldsymbol{\sigma}_h\} - \boldsymbol{\beta} \; \boldsymbol{\sigma}_h \; \right) d\Gamma + \int_{\Gamma_D} \boldsymbol{\sigma}_h \cdot \mathbf{n} v d\Gamma \\
J(u_h, v) &= -\sum_{e \in \Gamma_h \cup \Gamma_D} \int_e \frac{\eta}{h_e} u_h \cdot v \; d\Gamma
\end{aligned}
\tag{3.10}
$$

The symmetry of the FPM-Primal and the FPM-Mixed is reflected directly in Eqs. (3.8) and (3.9) when test and trial functions employ the same shape functions. Besides, we can see that Dirichlet and Neuman boundary conditions are imposed weakly in both FPM methods.

### 3.3 Numerical implementation

In this part, we will discuss the numerical implementation of the FPM. We take the FPM-Primal for example. The FPM-Primal can be written in the following matrix form,

$$\mathbf{Ku} = \mathbf{f}$$

where $\mathbf{K}$ is the global stiffness matrix, $\mathbf{u}$ is the vector with nodal DOFs, $\mathbf{f}$ is the load vector.

Substituting the shape functions $\mathbf{B}$ for $\nabla u$ and $\nabla v$, $\mathbf{N}$ for $u_h$ and $v$ into Eq. (3.8), the Point Stiffness Matrix $\mathbf{K}_E$ and the boundary stiffness matrices $\mathbf{K}_h$, $\mathbf{K}_D$ can be obtained as:

$$\mathbf{K}_E = \int_E \mathbf{B}^{\mathrm{T}} \mathbf{B} d\Omega \qquad \text{where } E \in \Omega.$$

$$
\begin{aligned}
\mathbf{K}_h = & \frac{-1}{2} \int_e (\mathbf{B}_1^{\mathrm{T}} \mathbf{n}_1^{\mathrm{T}} \mathbf{N}_1 + \mathbf{N}_1^{\mathrm{T}} \mathbf{n}_1 \mathbf{B}_1) d\Gamma + \frac{\eta}{h_e} \int_e \mathbf{N}_1^{\mathrm{T}} \mathbf{N}_1 d\Gamma \\
& + \frac{-1}{2} \int_e (\mathbf{B}_2^{\mathrm{T}} \mathbf{n}_2^{\mathrm{T}} \mathbf{N}_2 + \mathbf{N}_2^{\mathrm{T}} \mathbf{n}_2 \mathbf{B}_2) d\Gamma + \frac{\eta}{h_e} \int_e \mathbf{N}_2^{\mathrm{T}} \mathbf{N}_2 d\Gamma \\
& + \frac{-1}{2} \int_e (\mathbf{B}_2^{\mathrm{T}} \mathbf{n}_1^{\mathrm{T}} \mathbf{N}_1 + \mathbf{N}_2^{\mathrm{T}} \mathbf{n}_2 \mathbf{B}_1) d\Gamma + \frac{\eta}{h_e} \int_e \mathbf{N}_1^{\mathrm{T}} \mathbf{N}_2 d\Gamma \\
& + \frac{-1}{2} \int_e (\mathbf{B}_1^{\mathrm{T}} \mathbf{n}_2^{\mathrm{T}} \mathbf{N}_2 + \mathbf{N}_1^{\mathrm{T}} \mathbf{n}_1 \mathbf{B}_2) d\Gamma + \frac{\eta}{h_e} \int_e \mathbf{N}_2^{\mathrm{T}} \mathbf{N}_1 d\Gamma
\end{aligned}
\qquad \text{where } e = \partial E_1 \bigcap E_2. \quad (3.11)
$$

$$\mathbf{K}_D = \frac{-1}{2} \int_e \left( \mathbf{N}^{\mathrm{T}} \mathbf{n} \mathbf{B} + \mathbf{B}^{\mathrm{T}} \mathbf{n}^{\mathrm{T}} \mathbf{N} \right) d\Gamma + \frac{\eta}{h_e} \int_e \mathbf{N}^{\mathrm{T}} \mathbf{N} d\Gamma \quad \text{where } e \in \Gamma_D.$$

The global stiffness matrix $\mathbf{K}$ in the Fragile Points Method is generated by assembling all the submatrices, which is the same procedure as in the Finite Element Method. We can see that the global stiffness matrix is symmetric, sparse and positive definitive.

Considering the FPM-Mixed approach, we transform Eqs. (3.9) and (3.10) into the following matrices form.

$$
\begin{bmatrix} \mathbf{A} & \mathbf{B} \\ \mathbf{B}^{\mathrm{T}} & \mathbf{J} \end{bmatrix} \begin{pmatrix} \boldsymbol{\sigma} \\ \mathbf{u} \end{pmatrix} = \begin{pmatrix} \mathbf{f}_{\boldsymbol{\sigma}} \\ \mathbf{f}_{\mathbf{u}} \end{pmatrix} \tag{3.12}
$$

Employing the shape functions $\mathbf{M}$ for $\boldsymbol{\sigma}$ and $\boldsymbol{\tau}$, $\mathbf{B}$ for $\nabla u$ and $\nabla v$, $\mathbf{N}$ for $u_h$ and $v$, submatrices $\mathbf{A}_E$, $\mathbf{B}_E$, $\mathbf{B}_h$, $\mathbf{B}_D$, $\mathbf{J}_h$, $\mathbf{J}_D$ can be obtained

$$\mathbf{A}_E = \int_E \mathbf{M}^{\mathrm{T}} \mathbf{M} d\Omega \qquad\qquad \text{where } E \in \Omega .$$

$$\mathbf{B}_E = \int_E \mathbf{M}^{\mathrm{T}} \mathbf{B} d\Omega \qquad\qquad \text{where } E \in \Omega .$$

$$\mathbf{B}_h = \int_e \mathbf{M}_1^{\mathrm{T}} \left( \frac{\mathbf{1}}{2} \mathbf{n}_1 \text{ - } \boldsymbol{\beta}_e \right)^{\mathrm{T}} \mathbf{N}_1 d\Gamma + \int_e \mathbf{M}_2^{\mathrm{T}} \left( \frac{\mathbf{1}}{2} \mathbf{n}_2 \text{ - } \boldsymbol{\beta}_e \right)^{\mathrm{T}} \mathbf{N}_2 d\Gamma$$

$$+ \int_e \mathbf{M}_2^{\mathrm{T}} \left( \frac{\mathbf{1}}{2} \mathbf{n}_1 \text{ - } \boldsymbol{\beta}_e \right)^{\mathrm{T}} \mathbf{N}_1 d\Gamma + \int_e \mathbf{M}_1^{\mathrm{T}} \left( \frac{\mathbf{1}}{2} \mathbf{n}_2 \text{ - } \boldsymbol{\beta}_e \right)^{\mathrm{T}} \mathbf{N}_2 d\Gamma$$

$$\text{where } e = \partial E_1 \bigcap \partial E_2 . \qquad (3.13)$$

$$\mathbf{B}_D = \int_e \mathbf{M}^{\mathrm{T}} \mathbf{n}^{\mathrm{T}} \mathbf{N} d\Gamma \qquad\qquad \text{where } e \in \Gamma_D .$$

$$\mathbf{J}_h = \frac{\eta}{h_e} \int_e \mathbf{N}_1^{\mathrm{T}} \mathbf{N}_2 + \mathbf{N}_2^{\mathrm{T}} \mathbf{N}_1 - \mathbf{N}_1^{\mathrm{T}} \mathbf{N}_1 - \mathbf{N}_2^{\mathrm{T}} \mathbf{N}_2 d\Gamma \qquad \text{where } e = \partial E_1 \bigcap \partial E_2 .$$

$$\mathbf{J}_D = -\frac{\eta}{h_e} \int_e \mathbf{N}^{\mathrm{T}} \mathbf{N} d\Gamma \qquad\qquad \text{where } e \in \Gamma_D .$$

The global matrices $\mathbf{A}, \mathbf{B}, \mathbf{J}$ are also obtained by assembling the related submatrices.

If we employ constant functions for $\boldsymbol{\sigma}_h$ and linear functions for $u_h$, $\mathbf{B}$ and $\mathbf{M}$ are constant matrices, while $\mathbf{N}$ is linear then. The computation of integrals in submatrices $\mathbf{A}_E$, $\mathbf{B}_E$, $\mathbf{K}_E$ can be simply carried out just multiplying the integrand by the area of the corresponding subdomain. As for the integration on boundaries, Gauss integration method and direct analytic computation are both available. In this paper, we used Gauss integration method and employed two integration points for each boundary.

### 3.4 The way to model cracks in the present FPM

Reviewing the process of constructing trial and test functions through the Generalized Finite Difference Method, which leads to the discontinuity of trial functions, it will not cost much effort to introduce a crack or rupture between two subdomains. For example, if a crack appears at $\Gamma_{01}$ (shown in Figure 6), it changes from an internal boundary to a free boundary. Reviewing Eq. (3.8) of the FPM-Primal, we can find that we only need to delete those parts about $\Gamma_{01}$ on the left side, in other words, adjust the stiffness matrix **K** slightly and the right side stays the same. Therefore, there is no change of the number of the degrees of freedom and the dimensions of the global stiffness matrix and the load vector stay the same. This is simpler than remeshing, using Cohesive Zone Models, or Vanishing Element Methods, to model rupture and fracture. For example, in the work of Camacho and Ortiz (Camacho & Ortiz, 1996), the creation of cracks involves the modification of the boundary edge and node information, separating the neighboring elements by inserting new nodes. So, the number of degrees of freedom and the element connectivity are adjusted in every calculation step, which leads to the changes of the dimensions of the global stiffness matrix and the load vector.

The specific simulation of cracks and rupture by the FPM will be studied in our future work.

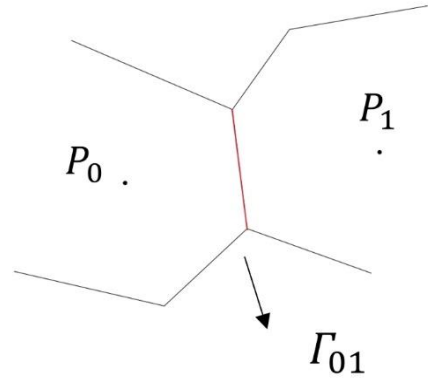

Figure 6. The boundary $\Gamma_{01}$

## 4. Numerical examples

In this section, different kinds of numerical examples are presented to demonstrate the consistency, robustness, high accuracy and convergence of the FPM. Because the results computed by the FPM-Primal method were found to be almost the same as those obtained from the FPM-Mixed method, we only present the solutions from the FPM-Primal in this paper.

For all the numerical examples in this section, linear trial and test functions are employed. The penalty coefficient $\eta = 10$ for the FPM-Primal in 1D, and $\eta = 5$ in 2D.

## 4.1 Patch test

To verify the consistency of the present FPM, we prescribe that the source function $f = 0$ in Eq. (2.1) and mixed boundary conditions are imposed. For the 1D patch test, the exact solution $u = x$, and for the 2D patch test, $u = x + y$.

Since both the exact solutions are linear functions, in order to pass these patch tests, numerical solutions of $u$ and its derivatives should be equal to the exact solutions.

In the 1D situation, when 11 Points are distributed either uniformly or randomly, errors of $u$ and $\dfrac{du}{dx}$ are less than $5 \times 10^{-14}$. In the 2D situation, with 121 Points distributed either uniformly or randomly, errors are less than $5 \times 10^{-14}$. The distributions of random points in 1D and 2D are shown in Figure 7(a) and (b), respectively.

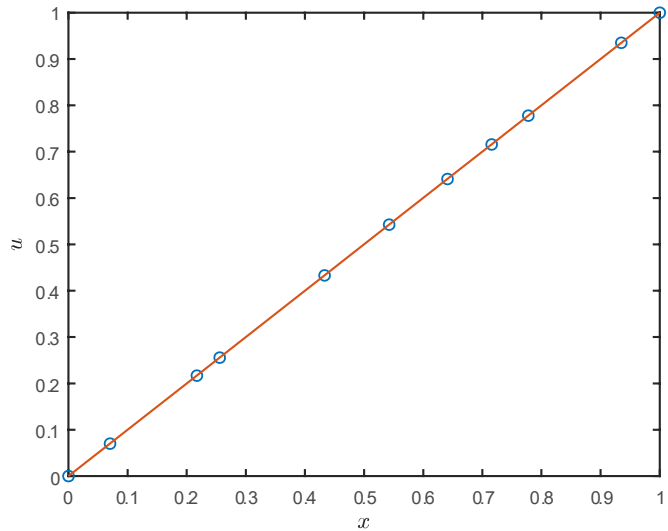

(a)

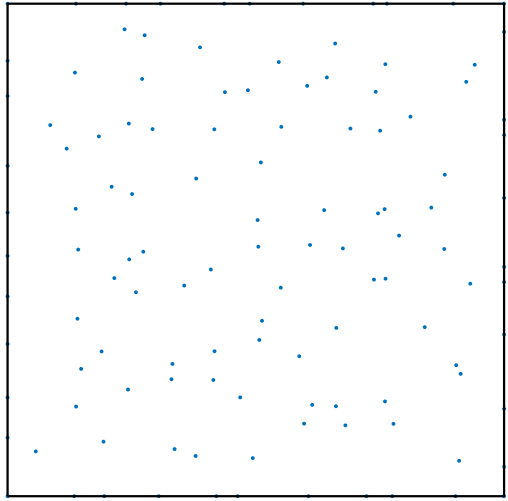

(b)

Figure 7 (a). The distribution of random points in 1D

(b). The distribution of random points in 2D

We can conclude that the present FPM is accurate enough in both 1D and 2D to pass these patch tests.

## 4.2 Poisson equation in 1D

In this part, we use the FPM to solve a 1D Poisson equation which is considered to be,

$$\frac{d^2u}{dx^2} = -1 \qquad x \in (0,1)$$

$$u_{x=1} = 0, \ u_{x=0} = 0$$

(4.1)

which corresponds to the postulated exact solution: $u = -\frac{1}{2}x(x-1)$.

When 30 Points are distributed either uniformly or randomly, numerical solutions of $u$ are shown in Figure 8(a) with the exact solution for comparison. The results of $\frac{du}{dx}$ are presented in Figure 8(b).

It can be seen that the FPM can yield highly accurate solutions when points are distributed uniformly and randomly. We can conclude that the FPM, with its Point Stiffnesses and Numerical Flux Corrections is a very robust approach.

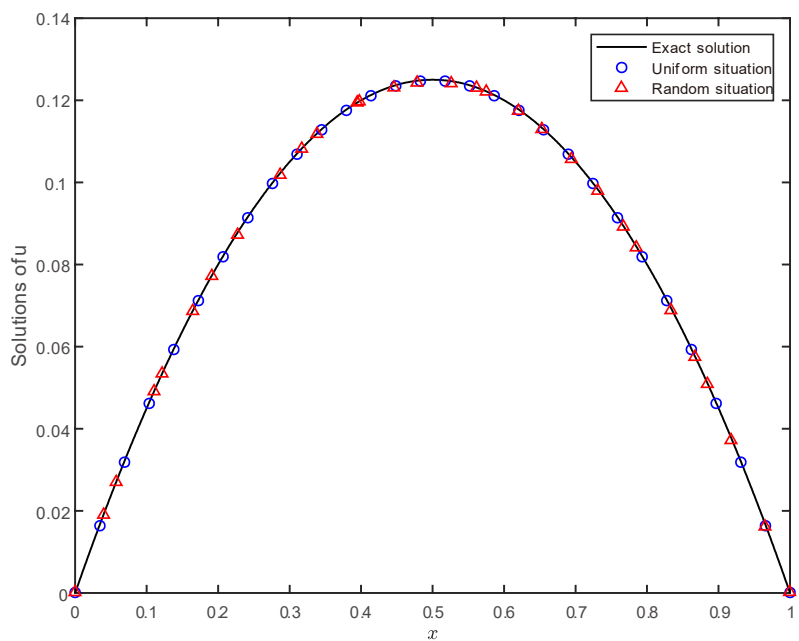

(a)

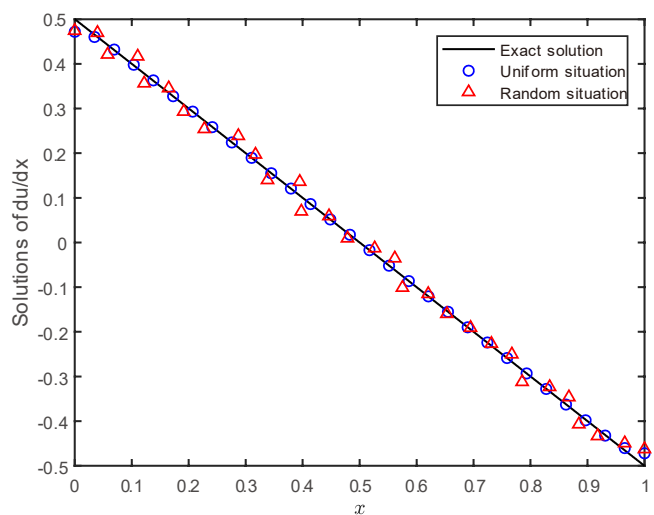

(b)

Figure 8 (a). Numerical solutions of $u$

(b). Numerical solutions of $\dfrac{du}{dx}$

## 4.3 Poisson equation in 2D

In this subsection, we continue to discuss the FPM for a 2D Poisson equation, which is defined in Eq. (4.2). The exact solution is postulated as $u = x(x-1)y(y-1)$.

Thus, the problem to be solved through FPM is taken to be:

$$\frac{\partial^2 u}{\partial x^2} + \frac{\partial^2 u}{\partial y^2} = 2\left(y^2 - y + x^2 - x\right) \quad (x, y) \in \Omega$$

$$\nabla u \cdot \mathbf{n} = y\left(y - 1\right) \quad (x, y) \in \Gamma_N \tag{4.2}$$

$$u = x\left(x - 1\right) y\left(y - 1\right) \quad (x, y) \in \Gamma_D$$

where $\Omega = \left\{ (x, y) \middle| \sqrt{x^2 + \left(y - 0.5\right)^2} \le 0.5, \ x \ge 0 \right\}$, which is a semicircle with the radius

equal to 0.5; Two types of boundaries

$\Gamma_D = \left\{ (x, y) \middle| \sqrt{x^2 + \left(y - 0.5\right)^2} = 0.5, \ (x, y) \in \Omega \right\}$, $\Gamma_N = \left\{ (x, y) \middle| x = 0, (x, y) \in \Omega \right\}$.

280 Points are distributed randomly in this semicircle. The Voronoi Diagram partition of the domain is shown in Figure 9.

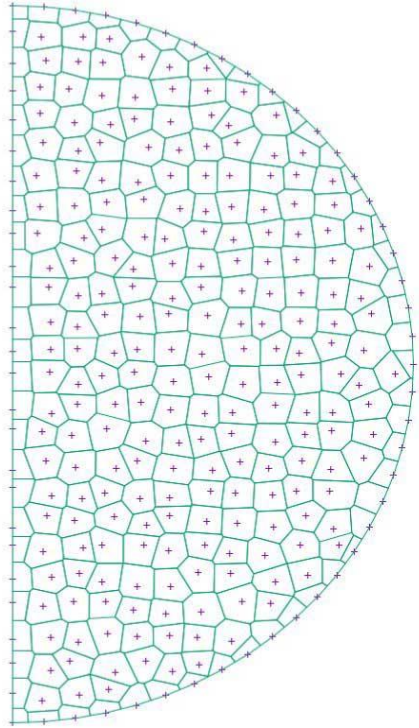

Figure 9. Partition of the domain into Points and Voronoi Diagram Partitions

The presently computed numerical solution of $u$ is shown in Figure 10(a), and the error of $u$ is shown in Figure 10(b).

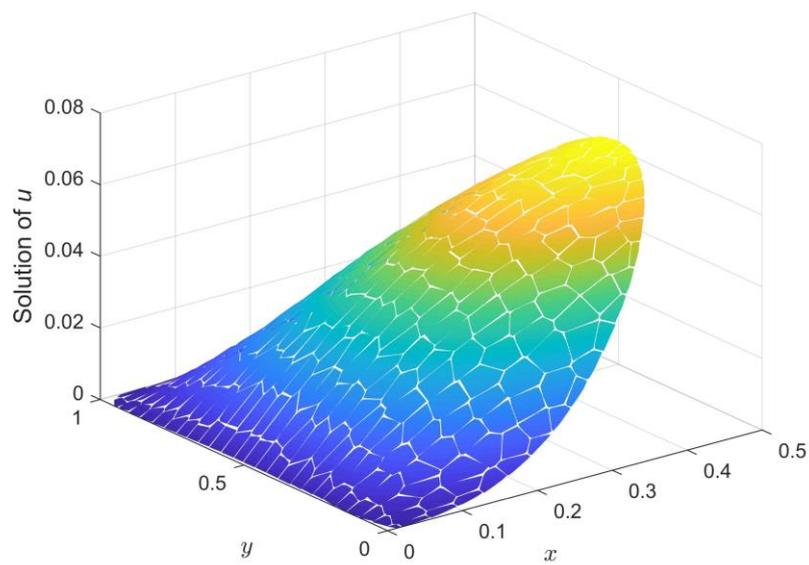

(a)

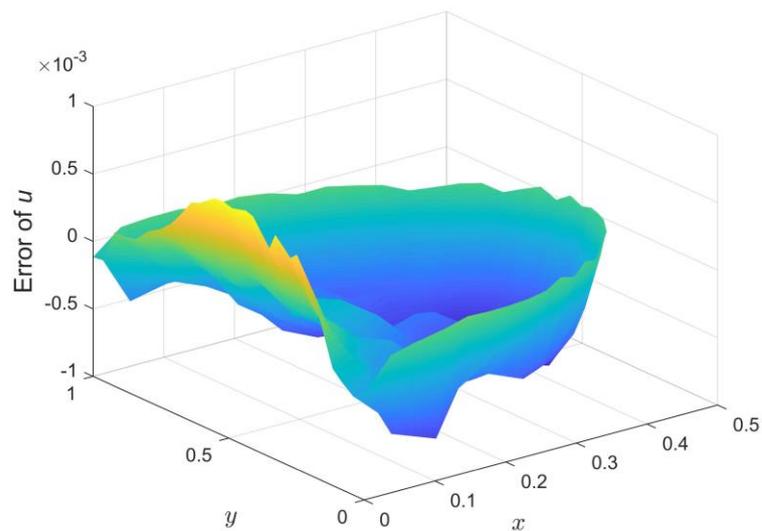

(b)

Figure 10 (a). The computed solution of  $u$

(b). The error of  $u$

## 4.4 The convergence study of the FPM

In order to estimate numerical errors clearly and accurately, we define the relative errors  $r_0$  and  $r_1$  with the Lebesgue norm.

$$r_0 = \frac{\left\| u^h - u \right\|_{L^2}}{\left\| u \right\|_{L^2}}$$

$$r_1 = \frac{\left\| \nabla u^h - \nabla u \right\|_{L^2}}{\left\| \nabla u \right\|_{L^2}}$$

where $\left\| u \right\|_{L^2} = \left( \int_\Omega u^2 d\Omega \right)^{\frac{1}{2}}$, $\left\| \nabla u \right\|_{L^2} = \left( \int_\Omega \left| \nabla u \right|^2 \right)^{\frac{1}{2}}$

For the 1D Poisson equation (shown in Eq. (4.3)), Dirichlet boundary conditions are imposed and the Points are distributed uniformly. The exact solution is taken as: $u = \frac{1}{2}(1-x)x$, $x \in (0,1)$. Thus, the problem to be solved through FPM is:

$$\begin{aligned} &\frac{d^2 u}{dx^2} = -1, \ x \in (0,1) \\ &u\big|_{x=0} = 0, \ u\big|_{x=1} = 0 \end{aligned} \tag{4.3}$$

The relations between $n$ (the number of Points) and the relative errors $r_0$ and $r_1$ are illustrated in Figure 11, where $R_0$ and $R_1$ stand for the convergence rates.

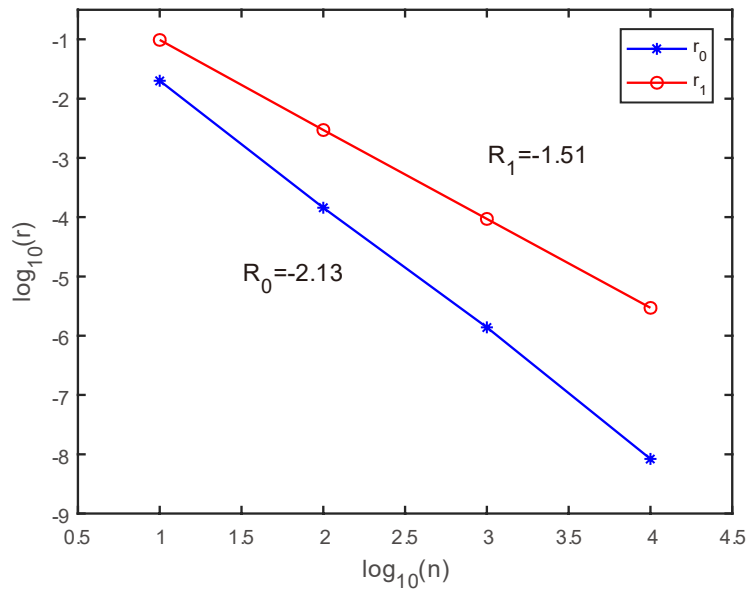

Figure 11. Relative errors for the 1D Poisson equation

Consider the 2D Poisson equation with Dirichlet boundary conditions (shown in Eq. (4.4)). Its exact solution is postulated as: $u = x(x-1)y(y-1)$, $x \in (0,1)$, $y \in (0,1)$. Thus, the problem to be solved through FPM is:

$$\frac{\partial^2 u}{\partial x^2} + \frac{\partial^2 u}{\partial y^2} = 2\left(y^2 - y + x^2 - x\right) \quad x, y \in (0,1)$$

$$u = 0 \quad \text{at} \quad x, y = 0, 1$$

$$(4.4)$$

In the domain, Points are distributed uniformly. The relations between $n$ and $r_0$, $r_1$ are shown in Figure 12.

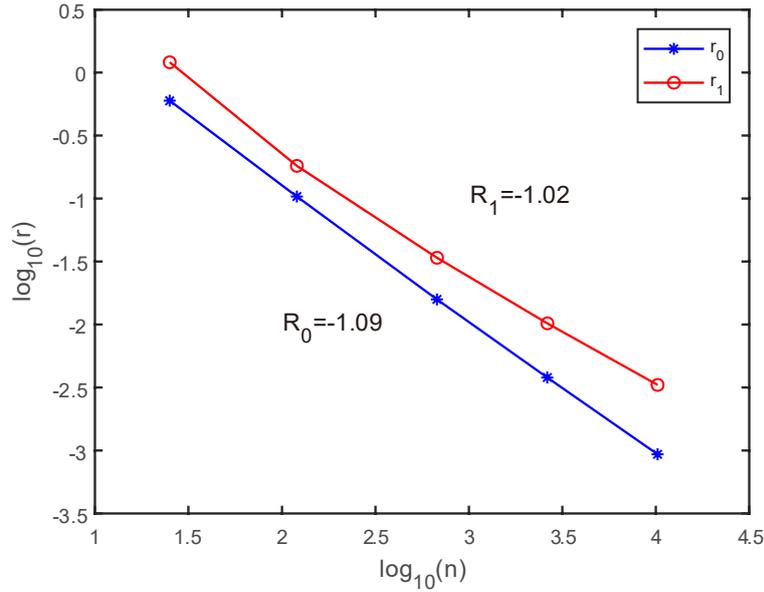

Figure 12. Relative errors for the 2D Poisson equation

From these two examples, the convergence and high accuracy of the FPM are apparent.

## 4.5 The influence of the penalty parameter

To study the influence of the penalty parameter, we take the 1D Poisson equation Eq. (4.3) and the 2D Poisson equation Eq (4.4) for instance. 101 and $101 \times 101$ points are distributed uniformly in 1D and 2D, respectively.

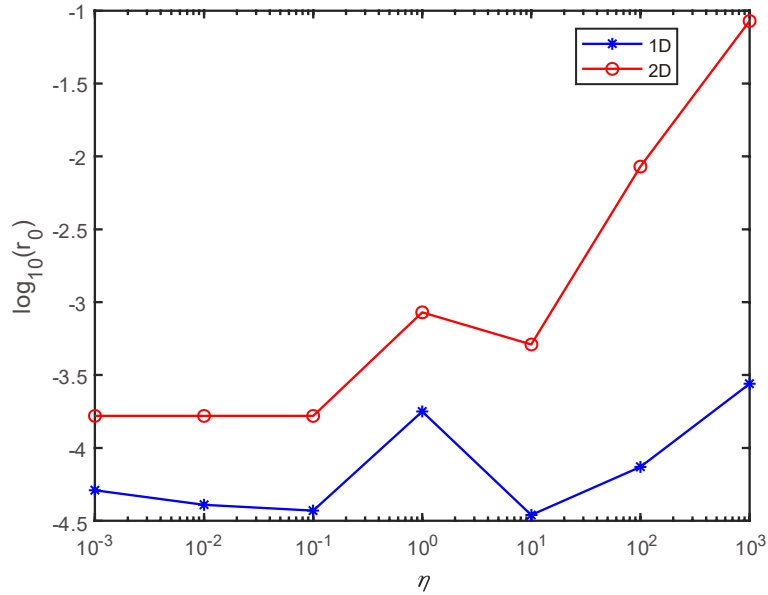

Figure 13. The relations between $\eta$ and $r_0$

In fact, a large penalty parameter yields a numerical solution with small jumps on the internal boundaries. Theoretically, if the penalty parameter is infinite, a continuous solution will be obtained. But the condition number of the global matrix will also increase when the penalty number becomes larger, which will influence the accuracy of the solution. The relation between the penalty parameter $\eta$ and the relative error $r_0$ is shown in the Figure 13. When $\eta > 100$, we can see the accuracy decreases when $\eta$ increases. Considering that the penalty parameter $\eta$ is required to be large enough to make the method stable, generally, the value of $\eta$ is usually chosen in the range of 1~100.

### 4.6 The comparison between FPM and FEM

For the FPM, we do not expect it to be much more accurate than the FEM under regular conditions. For example, a 1D Poisson equation with the exact solution: $u = (1-x)e^{-x^2}$ is solved by the FPM and FEM, respectively. Points or nodes are distributed uniformly. The relations between the number of Points and the relative errors are illustrated as below. It can be seen that the FPM is slightly more accurate in $\dfrac{du}{dx}$ but slightly less accurate in $u$.

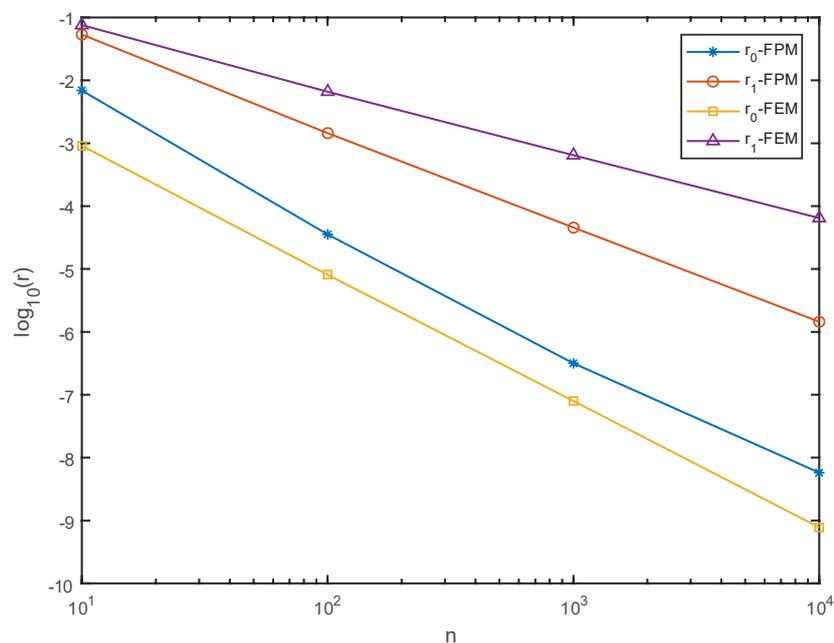

Figure 14. Relative errors of the FPM and FEM

However, we know that the FEM depends on a high-quality mesh, which makes it disadvantageous for modeling large deformation problems when elements become distorted. To illustrate the potential advantage of the FPM, we use the FPM to solve a 2D Poisson equation with the exact solution $u = x(x-1)y(y-1)$, $x \in (0,1)$, $y \in (0,1)$ where the distribution of 676 points is very irregular (shown in Figure 15). It can be seen that, the locations of some pairs of points almost coincide with each other, while some other pairs may have very large distances. Results show that the relative errors $r_0 = 0.018$ and $r_1 = 0.084$ for the FPM. Considering that the distribution of points is very irregular, the solution is quite acceptable. While the relative errors $r_0$ and $r_1$ derived from the FEM which employs linear triangle Elements based on the same randomly distributed Points are equal to 0.023 and 0.151, respectively.

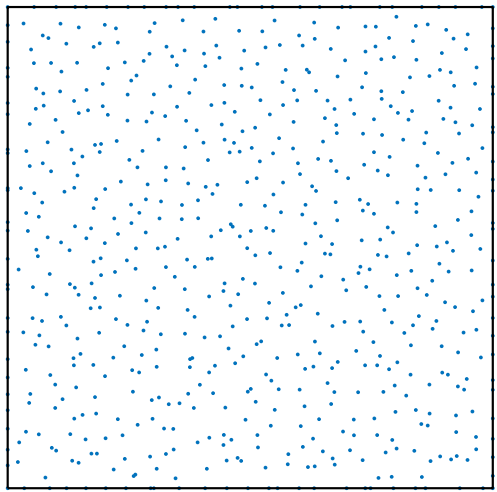

Figure 15. The distribution of random points in 2D

**Conclusion**

In this paper, a new Fragile Points Method (FPM) based on the concepts of Point Stiffnesses and Numerical Flux Corrections, is introduced. In the present FPM, local, very simple, discontinuous, polynomial, Point-based trial and test functions are employed. These functions are discontinuous in the overall domain. Thus, Numerical Flux Corrections are introduced. The symmetric and sparse global stiffness matrix is obtained as the summation of Point Stiffness Matrices. Several numerical examples are given to show the convergence, robustness, consistency and high accuracy of the FPM.

More importantly, as we mentioned in Section 3, the FPM possesses the ability to handle cracks through a very easy procedure, due to the discontinuity of trial and test functions. Besides, we can find that the present FPM can be easily parallelized. Therefore, it can be seen that the FPM has large potential for solving extreme problems. This will be demonstrated in our future studies.

**Acknowledge**

The first three authors thankfully acknowledge the support from the National Key Research and Development Program of China (No. 2017YFA0207800).